\documentclass[prl,superscriptaddress,twocolumn,floatfix]{revtex4}
\usepackage[latin1]{inputenc}  
\usepackage[T1]{fontenc}       
\usepackage[austrian,english,french]{babel}
\usepackage{graphicx}
\usepackage{amsmath}
\usepackage{amssymb}
\usepackage{amscd}
\usepackage[sort&compress]{natbib}

\newcommand{\ket}[1]{\ensuremath{\left| #1 \right\rangle}}

\begin{document}
\selectlanguage{english}

\title{Practical Quantum Key Distribution with Polarization Entangled Photons}
\author{A. Poppe}
\email{andreas.poppe@exp.univie.ac.at}
\author{A. Fedrizzi}
\affiliation{Institut für Experimentalphysik, Universität Wien,
Boltzmanngasse 5, 1090 Wien, Austria}
\author{T. Lorünser}
\author{O. Maurhardt}
\affiliation{ARC Seibersdorf Research GmbH (ARCS), 2444
Seibersdorf, Austria}
\author{R. Ursin}
\author{H. R. Böhm}
\affiliation{Institut für Experimentalphysik, Universität Wien,
Boltzmanngasse 5, 1090 Wien, Austria}
\author{M. Peev}
\author{M. Suda}
\affiliation{ARC Seibersdorf Research GmbH (ARCS), 2444
Seibersdorf, Austria}
\author{C. Kurtsiefer}
\author{H. Weinfurter}
\affiliation{Sektion Physik, Ludwig-Maximilians-Universität,
D-80797 München, Germany}
\author{T. Jennewein}
\affiliation{Institute for Quantum Optics and Quantum Information,
Austrian Academy of Sciences, Boltzmanngasse 3, 1090 Wien,
Austria}
\author{A. Zeilinger}
\affiliation{Institut für Experimentalphysik, Universität Wien,
Boltzmanngasse 5, 1090 Wien, Austria}
\affiliation{Institute for
Quantum Optics and Quantum Information, Austrian Academy of
Sciences, Boltzmanngasse 3, 1090 Wien, Austria}
\date{\today}

\begin{abstract}
We present an entangled-state quantum cryptography system that
operated for the first time in a real-world application scenario.
The full key generation protocol was performed in real-time
between two distributed embedded hardware devices, which were
connected by 1.45 km of optical fiber, installed for this
experiment in the Vienna sewage system. The generated quantum key
was immediately handed over and used by a secure communication
application. \vspace{0cm}
\end{abstract}
 \maketitle
\section{Introduction}
Quantum cryptography\,\cite{gisin02} is the first technology in
the area of quantum information that is in the process of making
the transition from purely scientific research to an industrial
application. In the last three years, several companies have
started developing quantum cryptography prototypes and the first
products hit the market \cite{idquantique,magiq,nec}.

Up to now, these commercial products are all based on various
implementations of the BB84\,\cite{bennett84} protocol. Because of
the unavailability of true single-photon sources, today's
commercially available quantum cryptography systems rely on
photons from attenuated laser pulses, as an approximation of the
single photon state. However, the produced state has a
non-vanishing probability to contain two or more photons per
pulse, leaving such systems prone to an eavesdropping through a
beam splitter attack. Although this attack is considered in recent
security proofs given for the BB84 protocol
\cite{luetkenhaus99,inamori01}, a true single photon source is
conceptually preferable.

\begin{figure}[h]
\center
\includegraphics[width=7 cm, keepaspectratio, clip=true, draft=false]{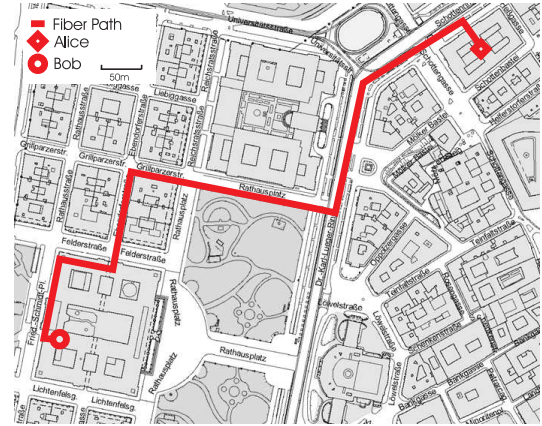}
\caption[City Map]{A quantum cryptography system is installed
between the headquarters of a large bank (Alice) and the Vienna
City Hall (Bob). The beeline distance between the two buildings is
about 650m. The optical fibers were installed some weeks before
the experiment in the Vienna sewage system and have a total length
of 1.45 km.} \label{fig:map}
\end{figure}

An elegant alternative is quantum cryptography based on entangled
photon pairs\,\cite{ekert91a}, which does not depend on photons
from attenuated laser pulses. In our implementation of
entangled-state quantum cryptography we make use of pairs of
single photons, which are individually completely unpolarized.
Information is only stored in correlations between the results of
measurements on the individual photons of a pair. With today's
entangled photon sources the probability of double pair emissions
is much lower than the probability of double photon emission in
weak coherent pules systems. Therefore entangled-state quantum
cryptography systems are closer to the original idea of quantum
cryptography. Furthermore, it can be shown \cite{luetkenhaus00}
that entanglement-based systems have a better performance in
certain situations. Additionally the randomness of the generated
key as required for a secure one-time-pad cipher\,\cite{vernam26}
is guaranteed by the quantum randomness of the measurement
process, whereas in attenuated laser quantum cryptography the
random encoding has to be achieved by using an external random
number generator. Thus, a fundamental point is that in
entangled-state quantum cryptography the key comes spontaneously
into existence at both measurement stations while in the BB84
protocol it still has to be transmitted from Alice to Bob.
Experimentally, entangled-state quantum cryptography has been
first demonstrated in 1998 using polarization-entangled photon
pairs\,\cite{jennewein00,naik00}. Alternative schemes are based
for example on energy-time entanglement \cite{tittel00}. Recent
development of compact, highly efficient sources for entangled
photons\,\cite{boehm03,aspelmeyer03b,trojek04} make
entangled-state quantum key distribution in real-life applications
feasible.

In this article we present an entangled-state quantum cryptography
prototype system in a typical application scenario. In the
experiment reported here, it was possible to distribute secure
quantum keys on demand between the headquarters of an Austrian
bank and the Vienna City Hall (see Figure \ref{fig:map}) using
polarization-entangled photon pairs. For key generation both
measurement stations switched randomly between complementary
bases\,\cite{wiesner83s} as also employed in the BB84 protocol.
The produced key was directly handed over to an application that
was used to send a quantum secured online wire transfer from the
City Hall to the headquarters of Bank-Austria
Creditanstalt\,\footnote{The first transfer of real money took
place on April 21st 2004.}.

\begin{figure*}[!]\vspace{0.1cm}
\includegraphics[width=17 cm, keepaspectratio, clip=true, draft=false]{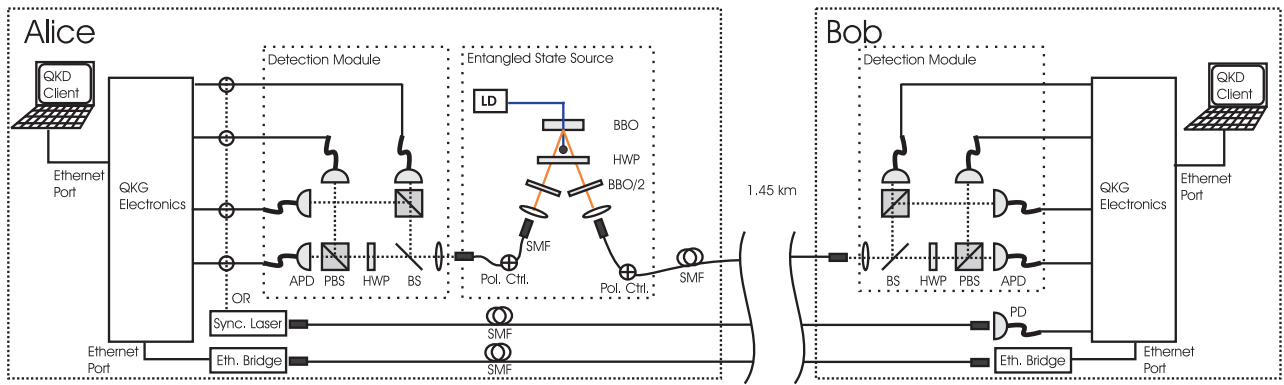}
\caption[Experimental Setup]{Sketch of the experimental setup of
the experiment. An entangled state source pumped by a violet laser
diode (LD) at 405nm produces polarization-entangled photon pairs.
One of the photons is locally analyzed in Alice's detection
module, while the other is sent over a $1.45$\,km long single-mode
optical fiber (SMF) to the remote site (Bob). Polarization
measurement is done randomly in one of two bases ($\ket{0°}$ and
$\ket{45°}$), by using a beam splitter (BS) which randomly sends
incident photons to one of two polarizing beam splitters (PBS).
One of the PBS is defined to measure in the $\ket{0°}$ basis, the
other is the $\ket{45°}$ basis turned by a half wave plate (HWP).
The final detection of the photons is done in passively quenched
silicon avalanche photodiodes (APD). When a photon is detected in
one of Alice's four avalanche photodiodes an optical trigger pulse
is created (Sync. Laser) and sent over a second fiber to establish
a common time basis. At both sites, the trigger pulses and the
detection events from the APDs are fed into an dedicated quantum
key generation (QKG) device for further processing. This QKG
electronic device is an embedded system, which is capable of
autonomously doing all necessary calculations for key generation.}
\label{fig:setup}
\end{figure*}

The quantum cryptography system used\,(see Figure \ref{fig:setup})
consists of the source for polarization-entangled photons located
at the bank, two combined polarization analysis and detection
modules and two electronic units for key generation. These two
quantum cryptography units which handled the five steps of secure
key generation - real-time data acquisition, key sifting, error
estimation, error correction and privacy amplification - are based
on an embedded electronic design and are compatible with classical
telecommunication equipment. The quantum channel between Alice and
Bob consisted of an optical fiber that has been installed between
the two experimental sites in the Vienna sewage system (by WKA /
CableRunner Inc.). The exposure of the fibers to realistic
environmental conditions such as stress and strain during
installation, as well as temperature changes were an important
feature of this experiment, as it shows that our system not only
works under laboratory conditions, but also in a realistic quantum
cryptography scenario.

\section{The Experimental Setup}
The entangled photon pairs used in our cryptography prototype were
produced using a compact device based on type-II spontaneous
parametric down-conversion\cite{kwiat95,kurtsiefer01}. Using a
16\,mW violet laser diode as pump source, it was possible to
generate and locally detect about $8.200$ pairs per second. The
photon pairs had a center wavelength of 810 nm and a FWHM
bandwidth of 5.6 nm. The visibility of the produced entangled
state $\ket{\psi^-}$ under local detection was above 96\%. For key
generation one photon of the pair was directly sent to Alice's
detection module (see Figure \ref{fig:setup}), while the other
photon was sent to the second receiver via $1.45$\,km of optical
fiber. The fiber used for the quantum channel was single mode for
810\,nm and had an average attenuation of about $3.2$\,dB per
kilometer resulting in a total attenuation of 6\,dB including the
connectors. The compensation of polarization rotation in the
fibers was done using fiber polarization controllers. Alignment of
the controllers was done offline before the quantum cryptography
protocol was started, using alignment software which calculated
and displayed the quantum bit error rate of the transmission
system in real-time based on the single photon detection events.
With two linear polarizers in both arms of the entanglement source
the polarization controllers were adjusted to compensate for the
arbitrary polarization rotation. It turned out that the
polarization was stable for several hours, therefore online
measurement and compensation for polarization drift was not
necessary.

A prototype of a currently developed dedicated QKD-hardware was
used for the first time in this experiment \cite{lieger04}. It
consists of three main computational components best suited to
manage all signal acquisition and QKD protocol tasks. All three
units are situated on a single printed circuit board. The main
task of the board is key acquisition, i.e. the processing of the
signals detected by the four detectors on each site of the QKD
system.

The board handles also the synchronization channel and generates a
strong laser pulse whenever a photon counting event is detected at
Alice's site. This is ensured by a logical OR connection of the
detector channels as shown on Figure 2. The synchronization laser
pulse is at the wavelength of 1550nm and it is sent over a
separate single mode fiber. The developed detection logic is
implemented in a FPGA and runs at a sampling frequency of 800MHz,
while employing a time window of 10 ns for matching the detection
events and synchronization signals. The synchronization mechanism
allows an identical realisation of the higher level electronics at
both sides and thus enables the development of universal QKD
devices.

The classical communication between the two devices is carried
over a TCP/IP connection, provided by an Ethernet bridge.

When fully developed the QKD hardware will be equipped with a full
scale QKD protocol modular library allowing seamless interchange
of relevant algorithms, as well as an encryption library,
comprising a number of state of the art encryption algorithms in
addition to the standard one-time pad. In addition to the main
"embedded" modus, the libraries can be used in a developer modus -
in the framework of an alternative computing environment (PC,
etc.).

In the current experiment the high speed electronics had only been
used for measuring detector events, hence, it worked in a
developer mode passing the data to an outside PC running the
protocol software stack and a graphical user interface (GUI). The
library modules utilized in the current experiment include data
acquisition, error estimation, error correction, implementing the
algorithm CASCADE \cite{brassard94}, privacy amplification and a
protocol authentication algorithm, ensuring the integrity of the
quantum channel\,\cite{peev04}, using a T\"oplitz matrix approach.
Furthermore the encryption-library modules applied include
one-time pad and AES encryption schemes, the latter allowing key
exchange on a scale determined by the user.

\begin{figure}[htbp]
\includegraphics[width=7 cm, keepaspectratio, clip=true, draft=false]{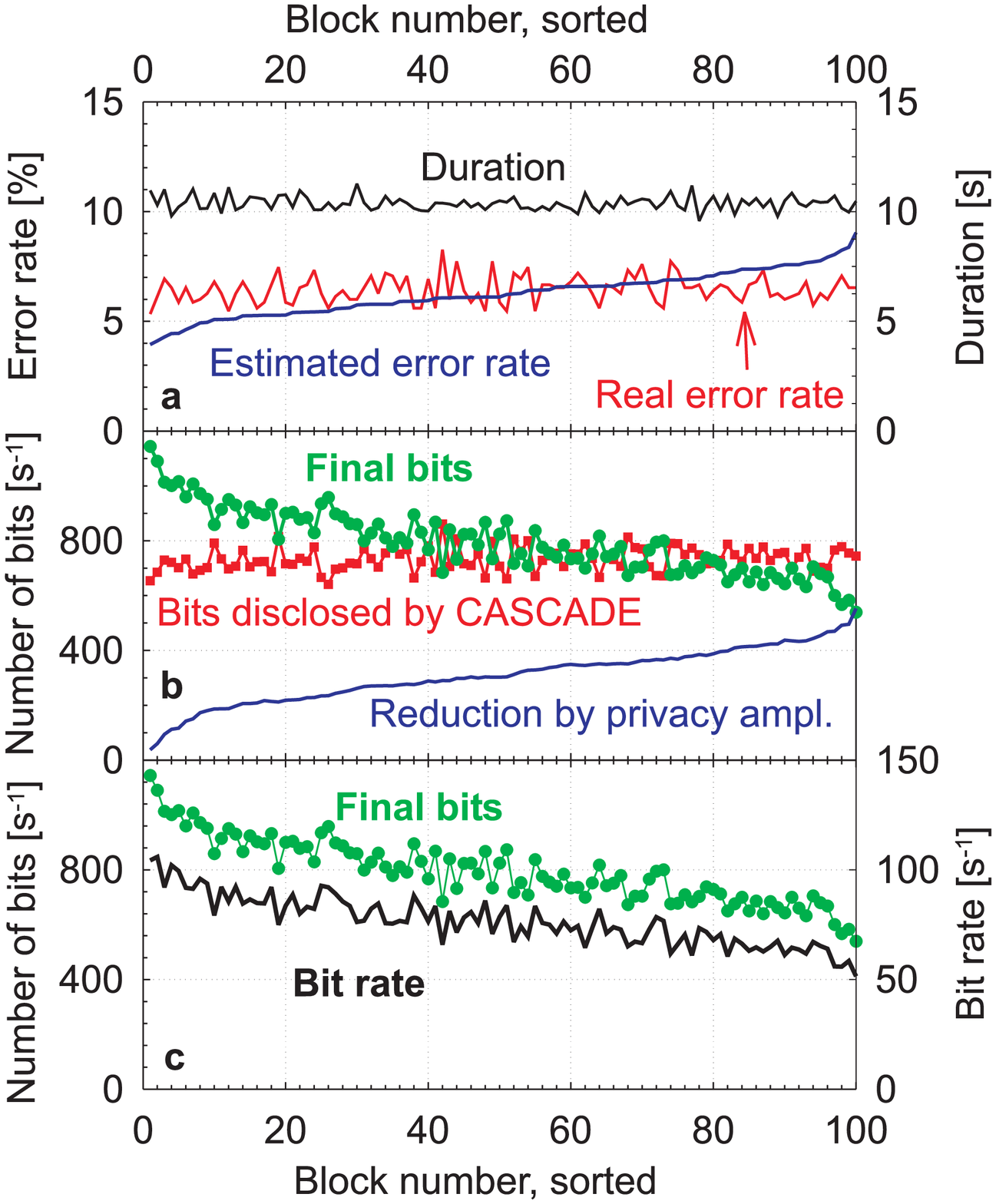}
\caption[Experimental Results]{Results obtained during 5 minutes
of the running experiment. The cryptography device worked with
blocks of raw data which were estimated to consist of 2500 bits
after sifting. Each key block was processed by the full quantum
cryptography software consisting of key sifting, error estimation,
error correction and privacy amplification (see text). The blocks
in this graph have been sorted by the QBER rather than the order
of their acquisition. \emph{(a)} In the topmost graph one can see
estimated QBER for the individual blocks as well as the real QBER
determined by directly comparing the whole sifted key over
subsequent measurement data blocks being processed by the device.
This calculation was only done for evaluation of the system and is
obviously not possible online for a real key exchange where it is
replaced by error estimation. Additionally one can see the time it
took to acquire the raw data of the given block. \emph{(b)}The
graph in the middle shows the length of the final key, the number
of bits disclosed by CASCADE and the number of bits discarded in
privacy amplification. \emph{(c)} In the lower graph one can see
the final secure bit rate produced by our system. }
\label{fig:results}
\end{figure}

The average total quantum bit error rate (QBER) was found to be
less than 8\,\%, for more than the entire run time of the
experiment. An analysis of the different contributions to the QBER
showed that about $2.5$\% comes from imperfections of the
detection modules, $1.2$\% from the imperfect production of the
entangled state. As these two contributions are practically not
accessible to eavesdropper\,\footnote{If the QBER contribution
related to our imperfect detection modules can be expressed as
unitary operations one could think of a method which would allow
an eavesdropper to reduce the QBER of our detection system in an
attack. However in a practical implementation this assumption is
unrealistic.} we subtracted these contributions for consideration
in privacy amplification. The rest of the QBER can be attributed
to the error produced by the quantum channel. For the data
presented in this article, the number of bits discarded in privacy
amplification is automatically adjusted to the number of bits
disclosed during error correction and the QBER of the quantum
channel. The average raw key bit rate in our system was found to
be about 80 bits/s after error correction and privacy
amplification. This value is mainly limited by the detection
efficiency of the avalanche photo diodes used as well as the
attenuation on the quantum channel.

\section{Secure Key Generation}
In this section we present a subset of the data obtained in our
experiment. It covers 18 minutes of data acquisition and key
generation. Within that time period Alice's detectors registered
overall more than 12.8 million counts. The same number of
synchronization pulses was sent to Bob. After determination of
coincidence events and key sifting a total of 244765 bits remained
for further processing by the classical algorithms. For this
purpose, the raw data was continuously grouped into blocks of
approximately 2500 bits, which where submitted individually to the
following classical protocols. Publicly announcing 25\% of the
bits in each block was done for estimating the quantum bit error
(QBER) of the sifted key. The average value of the estimated error
rate and the real error rate, calculated by directly comparing
Alice's and Bob's sifted key after the experiment, was found to be
$6.25$\% and $6.44$\% respectively. Even though the average value
of the estimated error rate matches the real error rate, the small
sampling size gives strong deviations in individual blocks.

The remaining 183574 bits after error estimation were then
submitted to the error correction process. Due to the disclosing
of bits during error correction, the useable key size has to be
further reduced to 110226. In the last stage of the protocol,
privacy amplification reduced the error-corrected key depending on
the estimated QBER for each individual block, leaving a total of
79426 bits for the final secure quantum key. This corresponds to
an average key distribution rate of 76\,bit/second.

Results for individual blocks are presented on Figure
\ref{fig:results}. There the blocks have been ordered by the
estimated quantum bit error rate rather than the order of their
acquisition. The acquisition time of the individual blocks,
plotted in the upper part of Figure \ref{fig:results}, is almost
constant and shows no correlation with the error rate. Also there
the estimated error rate is shown in comparison with the real
error determined after the experiment. One can see that the
estimated error nicely fits the real error rate. Only for a small
percentage of blocks, the error is over- or underestimated. By
increasing the size of the individual blocks and therefore
improving the sampling of the error estimation, this effect can be
mitigated. The over- and underestimation of the QBER can also be
seen in the second graph of Figure \ref{fig:results}, where the
amount of bits discarded by privacy amplification is plotted for
the individual blocks. In the last graph of Figure
\ref{fig:results} the bit rate of the individual blocks is shown.
One can clearly see the expected behavior: The number of final
bits is mainly dominated by the estimated error rate.

\section{Conclusion}
The experiment reported here demonstrates the operation of an
entangled-state quantum cryptography prototype system. All
calculations were done in real-time on a distributed embedded
system. Furthermore we demonstrate the successful run of our
quantum cryptography system in a real world application scenario
outside ideal laboratory conditions. The results clearly show that
entangled photon systems provide a good alternative for weak
coherent pulse systems, with the additional benefit of rendering
the beamsplitter attack much less efficient and thus being closer
to the ideal BB84 quantum cryptography idea than systems based on
weak coherent pulses.

\section{Acknowledgements}
This work was supported by the Austrian Science Fund (FWF), the
QuComm and RAMBOQ project of the European Commission, the
PRODEQUAC project  of the Austrian Ministry of Transportation
Innovation and Technology (FIT-IT Embedded Systems Program), the
City of Vienna (MA30), WKA and the Austrian BA-CA Bank. Special
thanks for strong support go to Helmut Kadrnoska and Richard
Plank.

\bibliographystyle{unsrt}
\bibliography{myLaTeX/physics,crypto}

\end{document}